# Two chemically similar stellar overdensities on opposite sides of the plane of the Galactic disk

Maria Bergemann[1] and Branimir Sesar[2] and Judith G. Cohen[3] and Aldo M. Serenelli[4,5] and Allyson Sheffield[6] and Ting S. Li[7] and Luca Casagrande[8,9] and Kathryn V. Johnston[10] and Chervin F.P. Laporte[10] and Adrian M. Price-Whelan[11] and Ralph Schönrich[12] and Andrew Gould[1,13,14]

[1] *Max Planck Institute for Astronomy, Koenigstuhl 17, 69117, Heidelberg, Germany*

[2] *Deutsche Börse AG, Mergenthalerallee 61, 65760 Eschborn, Germany*

[3] *Division of Physics, Mathematics, and Astronomy, California Institute of Technology, Pasadena, California 91125, USA*

[4] *Institute of Space Sciences (ICE, CSIC), Carrer de Can Magrans s/n, E-08193, Barcelona, Spain*

[5] *Institut d'Estudis Espacials de Catalunya (IEEC), C/Gran Capita, 2-4, E-08034, Barcelona, Spain*

[6] *Department of Natural Sciences, LaGuardia Community College, City University of New York, 31-10 Thomson Ave., Long Island City, NY 11101, USA*

[7] *Fermi National Accelerator Laboratory, P. O. Box 500, Batavia, IL 60510, USA*

[8] *Research School of Astronomy & Astrophysics, Mount Stromlo Observatory, The Australian National University, ACT 2611, Australia*

[9] *ARC Centre of Excellence for All Sky Astrophysics in 3 Dimensions (ASTRO 3D)*

[10] *Department of Astronomy, Columbia University, 550 W 120th st., Mail Code 5246, New York, NY, 10027 USA*

[11] *Department of Astrophysical Sciences, Princeton University, 4 Ivy Lane, Princeton, NJ 08544, USA*

[12] *Rudolf-Peierls Centre for Theoretical Physics, University of Oxford, 1 Keble Road, OX1 3NP, Oxford, United Kingdom*

[13] *Korea Astronomy and Space Science Institute, Daejon 34055, Korea*

[14] *Department of Astronomy, Ohio State University, 140 W. 18th Ave., Columbus, OH 43210, USA*

**Our Galaxy is thought to have undergone an active evolutionary history dominated by star formation, the accretion of cold gas, and, in particular, mergers up to 10 gigayear ago[1,2]. The stellar halo reveals rich fossil evidence of these interactions in the form of stellar streams, substructures, and chemically distinct stellar components[3,4,5].**

**The impact of dwarf galaxy mergers on the content and morphology of the Galactic disk is still being explored. Recent studies have identified kinematically distinct stellar substructures and moving groups, which may have extragalactic origin[6,7]. However, there is mounting evidence that stellar overdensities at the outer disk/halo interface could have been caused by the interaction of a dwarf galaxy with the disk[8,9,10].**

**Here we report detailed spectroscopic analysis of 14 stars drawn from two stellar overdensities, each lying about 5 kiloparsecs above or below the Galactic plane – locations suggestive of association with the stellar halo. However, we find that the chemical compositions of these stars are almost identical, both within and between these groups, and closely match the abundance patterns of the Milky Way disk stars. This study hence provides compelling evidence that these stars originate**

**from the disk and the overdensities they are part of were created by tidal interactions of the disk with passing or merging dwarf galaxies[11,12].**

We present the spectroscopic analysis of 14 stars from two diffuse structures in the Milky Way halo, separated vertically by more than 10 kpc: the Triangulum-Andromeda (TriAnd) and A13 overdensities[13,14,15,16,17]. TriAnd and A13 are located towards the Galactic anti-centre, at latitudes $-35$ deg < b < $-15$ deg and +25 < b < +40 deg. The age of stars in TriAnd is estimated at 6-10 Gyr from the colour-magnitude diagram[15]. Studies of motions of stars in these two structures revealed that they are kinematically associated[17] and could be related to the Monoceros Ring, a ring-like stellar structure that twists around the Galaxy. However, the nature of the TriAnd and A13 structures remains hotly contested, with formation scenarios ranging from a disrupted dwarf galaxy to their origin in the Galactic disk[18].

We obtained high-resolution spectra of 14 stars using the Keck and VLT telescopes (Extended Data Table 1). The stars are confirmed members of the A13 and TriAnd overdensities based on their radial velocities, proper motions, and photometry. We determine fundamental atmospheric parameters of the stars, as well as chemical abundances for O, Na, Mg, Ti, Fe, Ba, and Eu, by combining analysis of their colours with standard spectroscopic methods (Methods). We derive stellar distances from 2MASS photometry and spectroscopic gravities, which place the TriAnd stars at a Galactocentric distance of $r_{GC} = 18 \pm 2$ kpc (where 2 kpc ~ 1σ, one standard deviation, s.d.), roughly 5 kpc below the plane, and A13 at $r_{GC} = 16 \pm 1$ kpc, roughly 4 kpc above the Galactic disk plane (Figure 1). The typical distance uncertainties are ~ $1 - 2$ kpc (Extended Data Table 2). From the same spectra, we determine the line-of-sight velocity dispersion corrected for the solar motion and azimuthal velocity of stars (Methods), $\sigma_{los}$, to be 27 km s$^{-1}$, markedly smaller than that of the halo stars[12,19], which have $\sigma_{los} \approx 100$ km $s^{-1}$. The rotational velocity for the stars in the sample is 195 ± 25 km $s^{-1}$ consistent with the circular velocity in the outer disk.

Our analysis shows that the abundance distribution in A13 and TriAnd is extremely compact, and the spread is consistent with observational errors, which are of the order 0.15 dex. For the abundances, we use the notation [A/B], which refers to the logarithm of the abundance ratio of the chemical element A to the element B scaled to the solar value. Our results are: ⟨[Fe/H]⟩ = −0.59 ± 0.12 dex, ⟨[O/Fe]⟩ = 0.24 ± 0.11 dex, ⟨[Na/Fe]⟩ = 0.09 ± 0.11 dex, ⟨[Mg/Fe]⟩ = 0.20 ± 0.03 dex, ⟨[Ti/Fe]⟩ = 0.08 ± 0.09 dex, ⟨[Ba/Fe]⟩ = 0.14 ± 0.13 dex, ⟨[Eu/Fe]⟩ = 0.20 ± 0.16 dex (Extended Data Table 3), where the spread is given by the sample s.d. We compare these abundance ratios with literature measurements of stars from the Galactic disk and halo, dwarf spheroidal (dSph) galaxies, and globular clusters, finding that our measurements are consistent with abundances in the "thin" disk, the younger component of the Milky Way, and are inconsistent with all other stellar populations (Figure 2, Extended Data Figure 1). All but one star from the two overdensities lie directly on the metal-poor end of the thin disk track, which represents stars in the outer disk of the Galaxy[20]. The only TriAnd star with slightly lower metallicity, [Fe/H] ≈ −0.9, resides on the canonical "thick disk" track with higher [Ba/Fe] and lower [Na/Fe]. The overdensity abundances are also consistent with those derived from stars in open clusters[21,22] and with Cepheids[23,24] at comparable Galactocentric distances, in the $r_{GC}$ range 10 to 20 kpc.

The similarity of abundances suggests that the TriAnd and A13 stars have a common origin. However, their origin in a star cluster is very unlikely, as stellar clusters are compact in physical space, in sharp contrast to the overdensity stars. A tidal disruption of a globular cluster, for example such as known for Palomar 5[25], can cause individual stars to be strewn over large distances in one direction on the sky. However, unlike the overdensities, the tidal tails of Palomar 5 are thin in the transverse directions to the resulting stellar stream. Also, the A13 and TriAnd stars do not exhibit the anti-correlation between sodium and oxygen abundances, which is found in almost all globular clusters in the Milky Way[26].

On the other hand, dSph galaxies, which can extend over several kpc, and can be disrupted to extend over tens of kpc, show a much larger scatter in abundance space (Figure 2, Extended Data Figure 1), which is thought to be due to multiple generations of star formation[27]. Also, in contrast to the stars in the overdensities, dSph galaxies are known to extend to very low, typically sub-solar, [O/Fe], [Mg/Fe], and [Na/Fe] ratios at [Fe/H] ≈ −0.5. Fornax, a dwarf galaxy most close to TriAnd in metallicity, has [Na/Fe] ≈ −0.6 dex, almost one order of magnitude lower than the relative Na abundance of the A13 overdensity stars. On the other hand, the relative Ba abundance of the Fornax stars is a factor of 10 higher than that of the A13 and TriAnd stars.

The key challenge to the origin of the A13 and TriAnd stars in the Milky Way disk, although strongly supported by the stellar chemical abundances and their motions, is that the stars are located very far away from the disk plane and at large Galactocentric distances, $r_{GC} > 15$ kpc. A plausible scenario that may explain our observations is related to a merger of a dwarf galaxy with the Milky Way disk. Simulations show that such mergers can trigger vertical oscillations and flaring in the pre-existing disk, which naturally explains the existence of stellar overdensities above and below the Galactic midplane[9,28]. To test this scenario, we compare the spatial locations of the TriAnd and A13 stars with the predictions of an N-body model, which follows the interaction of the Sagittarius (Sgr) dSph galaxy with an initially stable Galactic disk[29] (Figure 3). The initial dark halo mass for the Sgr dSph progenitor was $M_{200} \sim 10^{11}$ $M_{\odot}$, which, after 5.57 Gyr of evolution, was stripped down to a bound mass of $3 \times 10^9$ $M_{\odot}$. Initially, the disk in the region of A13 and TriAnd, corresponding to heliocentric distances of 8 kpc < $r_{\rm helio}$ < 22 kpc, is confined between −5 deg < b < +5 deg. However, during its interaction with the Milky Way, Sgr is able to excite the outer disk out to −30 deg < b < +30 deg, kicking material well above the mid-plane out into the regions of our TriAnd and A13 stars. The results from these simulations – also considering the timescales of the interaction of the Sgr dSph with the Milky Way, which vary between 5 − 9 Gyr – support our interpretation that these stars may have originated from the disk.

**Acknowledgements**

We thank Iskren Georgiev for the help with telluric correction of the UVES spectrum. A.M.S was supported from the grant ESP2015-66134-R (MINECO). K.V.J's contributions were supported by a grant from the National Science Foundation AST-1614743. L.C. gratefully acknowledges support from the Australian Research Council (grants DP150100250, FT160100402). M.B. acknowledges support by the Collaborative Research center SFB 881 (Heidelberg University, subproject A5) of the Deutsche Forschungsgemeinschaft (DFG, German Research Foundation). We thank Steven Majewski and Katia Cunha for interesting diskussions on the topic and Jo Bovy for the help with implementing the disk flare profile. We are indebted to Thomas Müller for his assistance with the final production-quality versions of all figures.

**Author information**

Reprints and permissions information is available at www.nature.com/reprints. The authors declare no competing financial interests. Correspondence and requests for materials should be addressed to M.B. (e-mail: bergemann@mpia-hd.mpg.de).

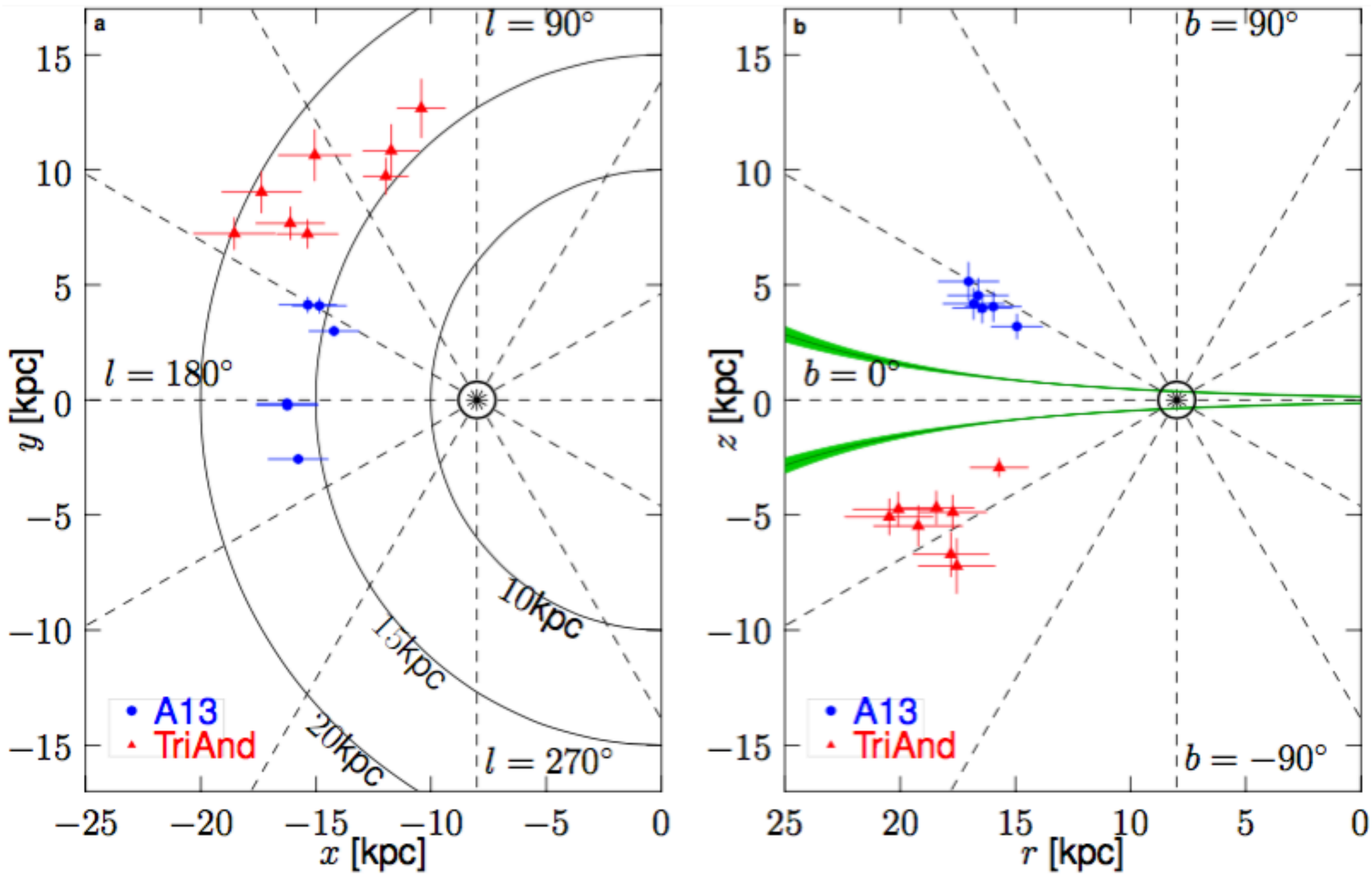

**Figure 1 |** The location of the observed stars in Galactocentric Cartesian coordinates.
The location of stars in the x-y and z-r plane, where x, y, and z are the directions in the Cartesian system of Galactic coordinates. $r_{GC}$ is the Galactocentric distance of the stars defined as $\sqrt{x^2 + y^2}$, in kiloparsec (kpc). The green curve in the right-hand-side panel indicates the flare profile of the Milky Way disk. The error bars represent the uncertainties of the distance measurements, which are estimated from the full posterior probability distributions as 1 s.d. confidence level.

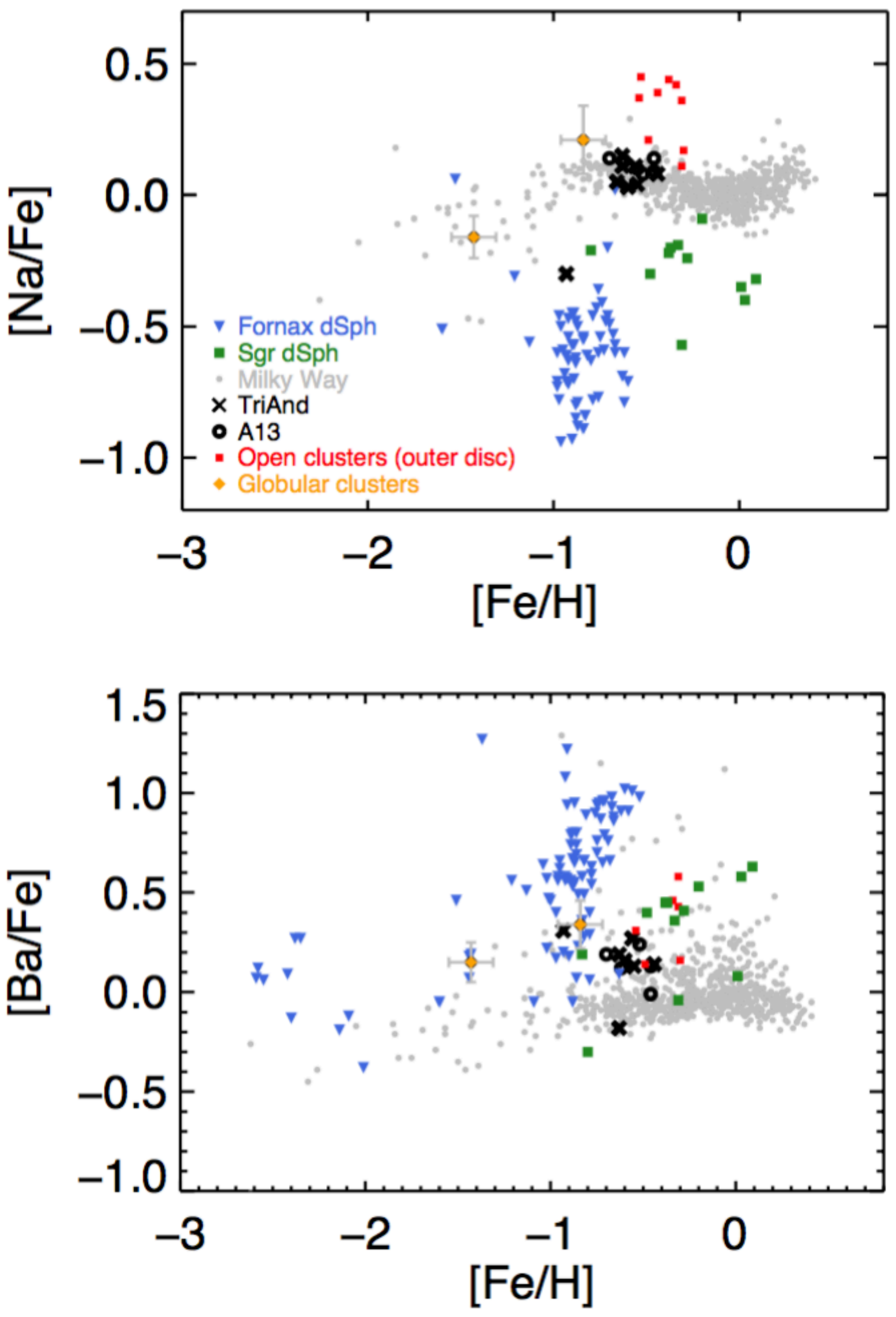

**Figure 2 |** Chemical abundances of the observed stars.
Chemical abundance ratios of [Na/Fe] and [Ba/Fe] against metallicity [Fe/H] in: the TriAnd (black crosses) and A13 overdensities (black circles), Milky Way disk and halo stars (grey circles), Fornax (blue symbols) and Sgr dwarf spheroidal galaxies (green symbols), globular clusters (orange symbols with error bars reflecting the intra-cluster abundance variation, derived as the root mean square variance (r.m.s.) of the sample, with N=13 for the 13 for M3 and N=25 for M71), and open clusters in the Galactic outer disk (red squares). The references are given in the Methods section. The typical uncertainty of the abundance measurements is 0.15 dex. The source data are provided in Extended Data Tables 1 and 3.

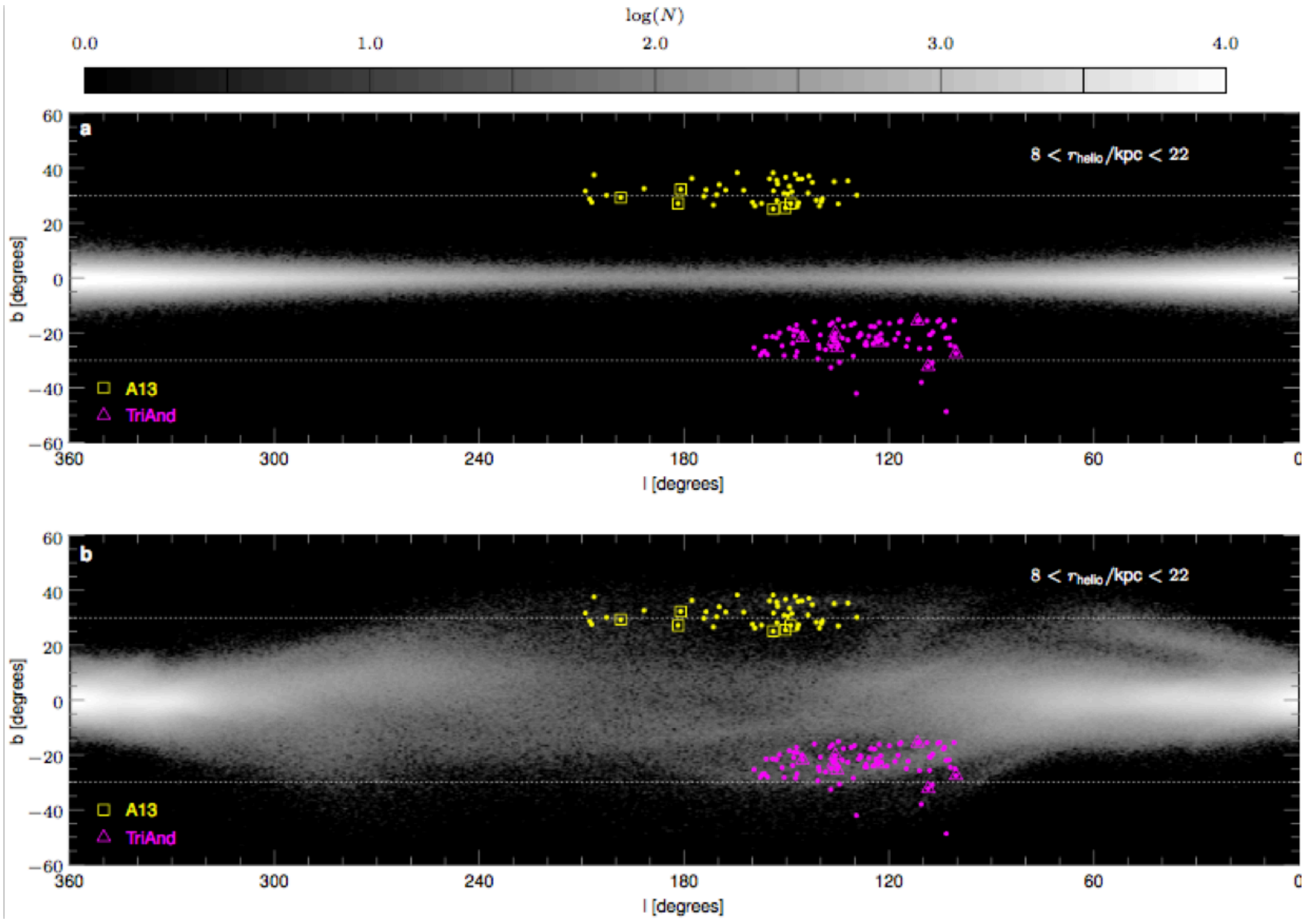

**Figure 3 |** Comparison of the observed positions of the observed stars with an N-body simulation of dwarf–Galaxy interaction.
The locations of the observed A13 and TriAnd stars (yellow squares and magenta triangles) compared with the two snapshots from the N-body model (represented as grey map), which follows the interaction of the Sagittarius dwarf spheroidal galaxy with an initially stable Galactic disk[29]. The top panel shows the distribution of star particles at the beginning of the simulation and bottom panel depicts the final distribution at present, after 5.6 Gyr. The confirmed members of the overdensities[15,17] are also shown with small yellow and magenta symbols.

## Methods

We acquired high-resolution spectra of 14 red giant branch (RGB) stars in the TriAnd and A13 overdensities[15,17]. The stars are confirmed members of the overdensities based on their photometry, radial velocities, and proper motions. We chose the brightest members of both groups in order to achieve the highest possible signal-to-noise within the available time on Keck. Metallicity was not a selection criterion. The photometric properties of the observed stars are reported in Extended Data Table 1. All observed stars are cool red giants on the upper part of the red giant branch (RGB) and their spectra are generally complex and display strong absorption features caused by molecules, circumstellar shells, and mass loss.

Thirteen stars were observed with the HIRES-R spectrograph (spectral resolution R of 36,000) at the Keck-1 telescope [30] and one star, TriAnd0_1, was observed using the UVES spectrograph (R of 47,000) at the VLT. The Keck spectra were taken on the night of Oct. 22, 2016, with typical exposure times of 20 to 30 minutes through thin cirrus. The UVES spectrum was taken on the night of Sept. 04, 2016 with 1 hour exposure. All Keck spectra cover the full optical region, from 4800 to 8770 Å, and the UVES spectrum covers the range from 4800 to 6800 Å. The signal-to-noise ratio (SNR/Å) of the HIRES spectra exceeds 200 near 5200 Å at the center of the echelle order. For the UVES spectrum, the average SNR/Å at the order center near 5500 Å is 50 and it increases to 85 at 6700 Å. The MAKEE pipeline that was specifically designed by T. Barlow was used to reduce HIRES spectra following standard procedures (i.e. bias subtraction, flat fielding, sky subtraction, order extraction, and wavelength calibration); the UVES spectrum was reduced using the ESO Reflex pipeline[31].

Detailed spectrum synthesis is essential to determine accurate chemical abundances. We use the MARCS[32] stellar model atmosphere grid, because it densely covers the parameter space of the stars we are interested in. Moreover, it also accounts for the necessary molecular opacities, such as MgH[33], that plague the spectra of cool RGB stars. The MARCS models are spherically-symmetric for low-gravity giants and line blanketing is treated using the opacity sampling method. The models account for the radiation pressure on molecules, although this does not affect stars in the temperature range and evolutionary stage in our sample.

Stellar atmospheric parameters are determined using several techniques. We attempt to follow our standard procedure[34] to derive the atmospheric parameters as closely as possible for the program RGB stars. For cool red giants, spectroscopic estimates based on the excitation-ionization equilibrium of iron provide reliable diagnostic of the gravity and metallicity. We used the Infrared Flux Method[35,36] to determine $T_{eff}$. Optical and infrared magnitudes were taken from the APASS and 2MASS photometry and corrected for interstellar reddening[37]. Surface gravities, metallicities, and microturbulent velocities were determined by means of excitation and ionization balance of Fe I and Fe II lines. NLTE corrections for the stars with log(g) ~ 1 and metallicity [Fe/H] above −1 dex are very small[38], and barely affect the estimates. The derived stellar parameters and their uncertainties are reported in Extended Data Table 2. The uncertainties of $T_{eff}$ were derived using the standard approach[36]. The uncertainties of log(g) and [Fe/H] are 0.15 and 0.1 dex, respectively. They represent the total uncertainty of the method, including the systematic and random error components[34].

Abundances are computed for the chemical elements O, Na, Mg, Ti, Fe, Ba, and Eu, using the least blended spectral lines that are detected in the observed spectra. Line fitting is done using spectral synthesis with the SME code[39]. For Mg, we use two diagnostic lines at 5528 and 5711 Å, adopting experimental data for atomic transitions[40]. For Mg, the mean NLTE abundance corrections are −0.10 for the 5711 Å line and 0.02 dex for the 5528 Å line[40]. The Fe linelist contains 123 lines of Fe I and Fe II[41]. Oxygen abundance was derived using the two forbidden [OI] lines at 6300 and 6363 Å, with log(*gf*) = −9.717 and −10.185 dex[42], respectively. In the UVES spectrum, the 6300 Å line is contaminated by telluric absorption lines, therefore the spectrum is first corrected using the ESO Molecfit package[43]. The 3D NLTE abundance corrections are negligible for the forbidden oxygen lines in our regime of stellar parameter space[44]. Na abundances are measured using the features at 5682 Å and 5688 Å (6 stars), or 6154 Å and 6160 Å (7 stars). This is because slightly different settings are used for Keck observations, and for a given setting one of the Na I doublets falls in the gap between spectral orders. However, all four Na I features are available in the UVES spectrum of TriAnd0_1, and they give consistent abundances. The 6154 Å and 6160 Å features are also used to estimate Na abundance in Fornax dSph galaxy[45]. The spectrum for one of the Keck targets is shown in Extended Data Figure 2, with prominent lines in the region around the 6154 Å Na I line labeled. The NLTE corrections for Na I lines are ≈ −0.12 dex for the 6154 and 6160 Å lines and ≈ −0.14 dex for the 5682 and 5688 Å lines[46]. For Ti, we used 23 lines, including 18 lines of Ti I and 5 lines of Ti II, which are least blended by molecular transitions, in particular, MgH, which is a major contaminating species at wavelengths below 6000 Å. We use LTE Ti abundances in this work, because NLTE Ti model does not give consistent solutions with 1D hydrostatic models[47]. Ba abundances are determined using the Ba II lines at 5853, 6141, and 6496 Å applying NLTE corrections[48]. The NLTE corrections for these lines are −0.03 to −0.05 dex. The isotopic shifts and hyperfine splitting (HFS) are also taken into account[49,50,51]. The only Eu II line that can be measured in the spectra is the feature at 6645 Å, which is affected by isotopic and HFS splitting. The main isotopes are $^{151}$Eu and $^{153}$Eu, with the solar abundances of 47.8 and 52.2 % respectively. The isotopic shifts, HFS magnetic dipole and electric quadrupole constants, are taken from experimental studies[52,53]. The average solar-scaled abundance ratios are reported in Extended Data Table 3.

Every effort has been made to check the accuracy of each abundance measurement; all spectral fits are examined by eye. To estimate the uncertainties in the chemical abundances, we follow our standard procedure[40]. The typical measurement error is ≈ 0.15 dex. The individual abundance errors are given in Extended Data Table 3. If the atomic lines of interest are contaminated by blends, the measurements have been deemed unreliable and no abundance is provided. Furthermore, we determine the solar abundances using the same linelist as for the program stars. Our derived NLTE solar abundances are in a very good agreement with reference estimates[54]. The stellar abundances are taken relative to our solar abundances.

The distances to TriAnd and A13 stars are estimated using the Bayesian method[55], with the 2MASS colours and the stellar parameters (Figure 2). The distances are the median of the posterior probability distribution (PDF) functions, and uncertainties are estimated from the full PDF as 1 s.d. confidence level. Our new distances are ~ 30% lower than those determined by our group previously, because that work used an approximate linear relationship between the absolute magnitude and colour of a star including a parameterised metallicity term. This revision is not crucial for this work, because it applies to all stars in the overdensities and does not affect the stellar membership classification. Figure 2 also shows the average disk scale height profile for the low-α stellar population[56].

The line-of-sight velocity dispersion was derived from the measured radial velocities (Extended Data Table 2), after correcting them for the Galactic standard of rest[57] and for the average motion of stars in the azimuthal direction (Extended Data Figure 3). The raw data represent the measured radial velocities, from which we first subtract the line-of-sight motion to the Sun, that is, we show the estimated line-of-sight velocity in the Galactic standard of rest, $V_{\mathrm{GSR}} = V_{\mathrm{LOS}} - V_{\mathrm{Sun}} \cdot (\bar{r} - \bar{r}_{\mathrm{Sun}})$, where $V_{\mathrm{Sun}}$ is the velocity of the Sun in the Galactic standard of rest[57], $\bar{r}$ is the position vector of the star, and $\bar{r}_{\mathrm{Sun}}$ is the position vector of the sun. In a second step, we correct for the projected rotation component of the stellar population, i.e. we define $V' = V_{\mathrm{GSR}} - V_{\mathrm{rot}} \cdot (\bar{r} - \bar{r}_{\mathrm{Sun}})$, where $V_{\mathrm{rot}}$ is an estimate for the average motion in the azimuthal direction.

The abundances in Figure 3 and in Extended Data Figure 1 are compared with the following data: Galactic disk and halo stars[20]; Sgr dSph galaxy[58,59,60] (ref. 60: only abundances with uncertainties of less than 0.1 dex), Fornax[45,61,62,63]; globular clusters (M3[64], M71[65]); open clusters Be 25 and NGC 2243 in the outer disk of the Milky Way[21,22]. The [Fe/H] and [Na/Fe] data for the Galactic disk were derived in NLTE. The [Ba/Fe], [Ti/Fe], and [Mg/Fe] data for the Galactic disk stars are LTE estimates. All literature data for the dSph systems, globular and open clusters, represent LTE estimates. The NLTE corrections tend to lower [Ba/Fe] and [Na/Fe] for red giants with sub-solar metallicity. For the dwarf stars, which constitute the comparison sample for the Milky Ways disk, the [Ba/Fe] NLTE corrections are negligible, within −0.02 dex. The typical NLTE corrections for the Mg I lines in the spectra of dwarfs are ~ 0.15 dex, and for RGB stars of the order −0.10 dex (5711 Å line) or close to zero (5528 Å line). Therefore, our conclusions would not be affected by the fact that some abundances were derived using LTE.

## Data Availability Statement

All data relevant to the manuscript are available from the authors. The N-body simulation data shown in Figure 3 are available by request. The data shown in Figure 1 and in Extended Data Figures 1, 2 are included with the manuscript as source data. The data shown in Figure 2 and in Extended Data Figure 3 are provided in Extended Data Tables 1 and 3.

The HIRES spectra are available at the Keck Observatory Archive, funded by NASA

https://www2.keck.hawaii.edu/koa/public/koa.php

The UVES data (Program ID: 097.B-0770(A)) are available from ESO Science Archive Facility at http://archive.eso.org/eso/eso_archive_main.html

The code used to determine stellar parameters and abundances is available at http://www.stsci.edu/~valenti/sme.html

The input linelist can be provided at the request to the corresponding author. The MARCS model atmospheres are available at the developer's website http://marcs.astro.uu.se

The NLTE abundance corrections for O, Mg, Ti, and Fe lines are available at the online database http://nlte.mpia.de

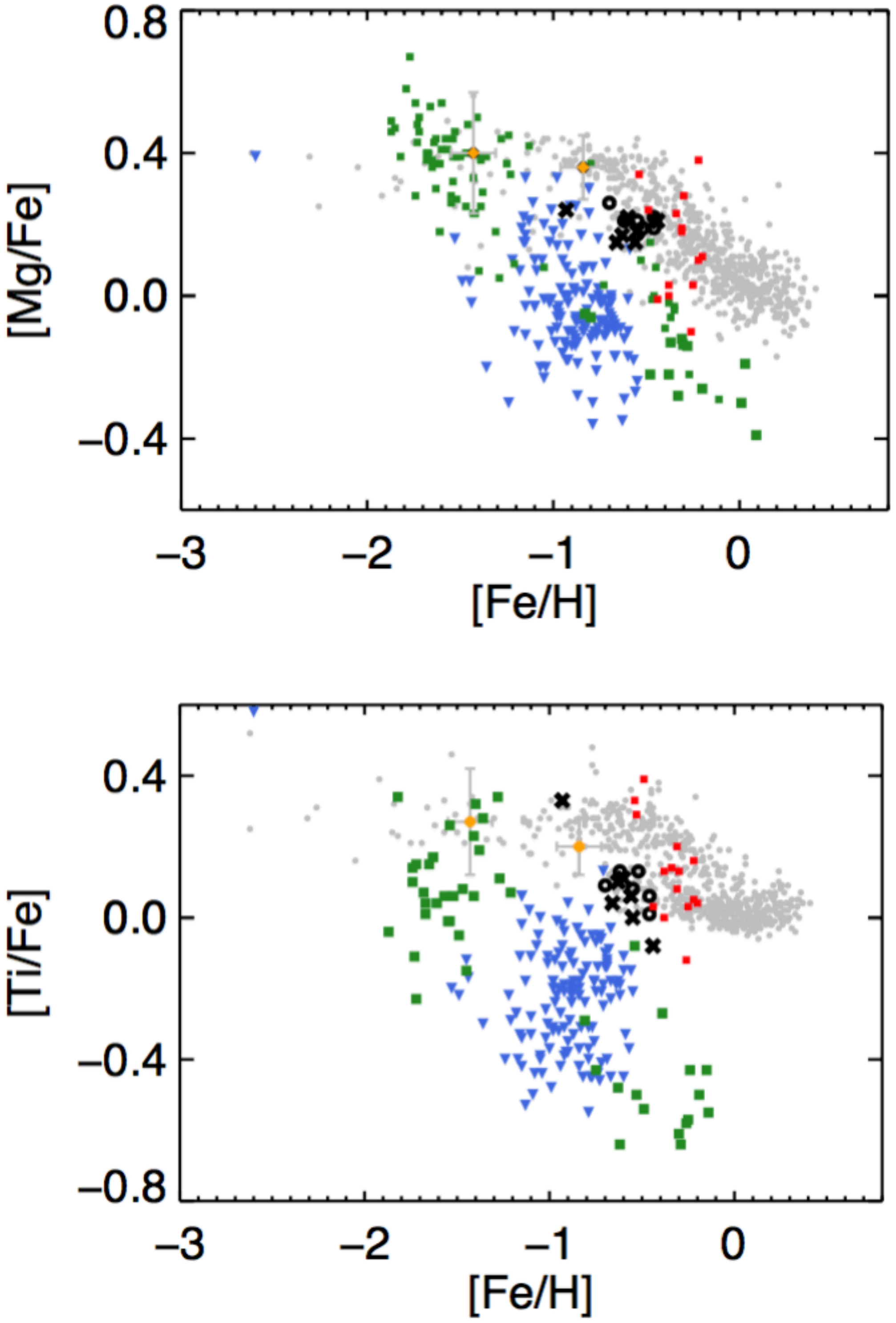

Extended Data Figure 1 **|** Chemical abundances of the observed stars. Chemical abundance ratios of [Mg/Fe] and [Ti/Fe] against metallicity [Fe/H] in the TriAnd (black crosses) and A13 overdensities (black circles). Symbols as in Figure 3. The source data are provided in Extended Data Tables 1 and 3.

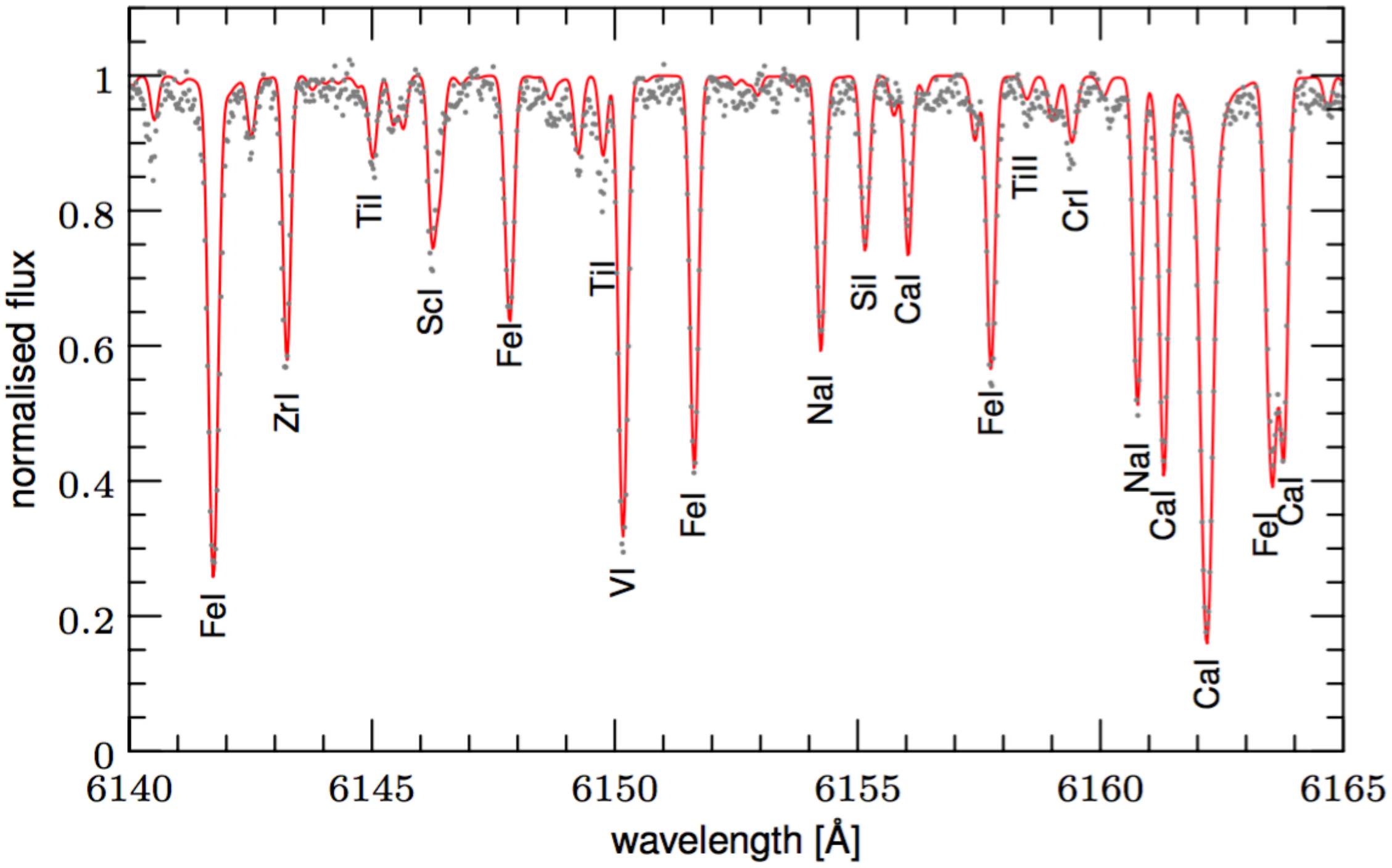

Extended Data Figure 2 **|** Comparison of the observed and a model spectrum for a star in the A13 overdensity.

The Keck spectrum of the star 2MASS 07154242+6704006 (black line) and the best-fit model spectrum (red line). The two Na I lines at 6154 and 6160 Å are used to determine Na abundance of the star.

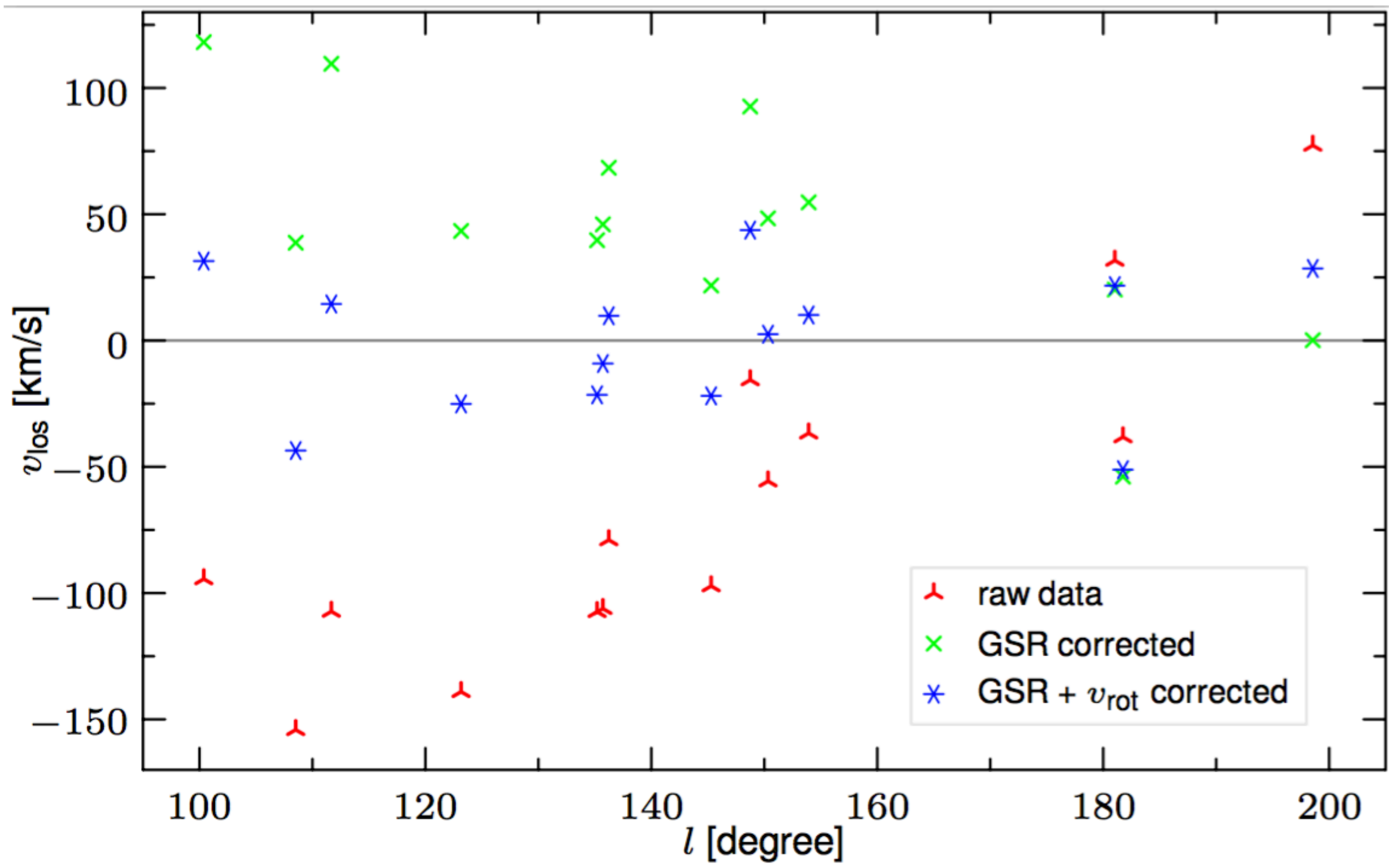

Extended Data Figure 3 | The line-of-sight velocities of the observed stars against galactic longitude *l*. See methods.